\documentclass[11pt,a4paper]{article}

\usepackage[utf8]{inputenc}
\usepackage{amsmath,amssymb}
\usepackage{graphicx}
\usepackage[colorlinks=true, linkcolor=blue, citecolor=blue, urlcolor=blue]{hyperref}
\usepackage{geometry}
\usepackage{authblk}
\usepackage{cite}
\usepackage{microtype}

\title{\textbf{Natural Quenching of Inner Viscous Torques in Accretion Disks Around Mini-Boson Stars}}

\author[1]{Sandip Dutta}
\affil[1]{Department of Applied Mathematics, Dinabandhu Andrews Institute of Technology and Management, Kolkata \protect\\ \texttt{duttasandip.mathematics@gmail.com}}
\date{}

\begin{document}
	
	\maketitle
	
	\begin{abstract}
		The standard Novikov-Thorne model for geometrically thin accretion disks assumes a zero-torque inner boundary condition at the innermost stable circular orbit (ISCO). Relaxing this assumption to allow a finite viscous torque---driven by magnetic stresses bridging the plunging region---can significantly enhance the radiative efficiency and high-frequency spectral emission of black hole accretion flows. While this mechanism is well-established for Kerr spacetimes, its thermodynamic viability around exotic, horizonless compact objects remains largely unexplored. In this article, we investigate the application of the inner viscous torque boundary condition to a thin disk embedded in a regular mini-boson star spacetime. By numerically integrating the coupled Einstein-Klein-Gordon field equations and employing the Lense-Thirring slow-rotation approximation, we generate an exact, horizonless background that strictly satisfies the weak energy condition. We demonstrate that the topological absence of a coordinate ISCO fundamentally alters the accretion dynamics. Without a sudden geodesic plunge, stable circular orbits penetrate deep into the dense scalar core, causing the physical inner boundary of the disk to be determined dynamically by the peak of the Keplerian angular velocity. Because the specific angular momentum of the fluid approaches zero at this deep microscopic boundary, the standard viscous torque is mathematically quenched. Consequently, the enhanced radiative flux divergence typically associated with boundary torques in black holes is naturally suppressed around regular boson stars. This torque quenching offers a robust phenomenological discriminator between standard black holes and horizonless ultralight scalar structures.
	\end{abstract}
	
	\vspace{2ex}
	
	\section{Introduction}
	
	The accretion of matter onto compact objects is one of the most efficient energy extraction mechanisms in the universe, powering active galactic nuclei (AGN), quasars, and X-ray binaries. For decades, the foundational framework for modeling these systems has been the geometrically thin, optically thick disk model pioneered by Shakura and Sunyaev \cite{Shakura1973}, and extended to general relativity by Novikov, Thorne, and Page (NTP) \cite{Novikov1973, Page1974}. A central tenet of the standard NTP model is the ``no-torque'' inner boundary condition. It assumes that as the accreting fluid crosses the innermost stable circular orbit (ISCO), it enters a rapid, ballistic plunge toward the event horizon, abruptly terminating the viscous transport of angular momentum and rendering the inner edge of the disk stress-free.
	
	However, subsequent theoretical and magnetohydrodynamic (MHD) studies have challenged the universality of the zero-torque assumption. Magnetic field lines can remain anchored in the plunging fluid or the rotating event horizon itself, transmitting significant torques back across the ISCO into the main body of the accretion disk \cite{Krolik1999, Gammie1999}. As demonstrated by Agol and Krolik \cite{Agol2000}, even a modest non-zero specific inner torque ($\mathcal{T}_{in}$) injects additional mechanical work into the inner disk regions. This boundary interaction sharply diverges the localized radiative flux, significantly elevates the global bolometric efficiency, and hardens the observed spectrum. Such enhancements are frequently invoked to resolve observational tensions, such as the extreme radiative efficiencies ($\eta_{bol} > 12\%$) required by luminous quasars \cite{Davis2011} and the ubiquitous soft X-ray excess observed in Seyfert galaxies \cite{Done2012}.
	
	Concurrently, the study of exotic compact objects (ECOs) has gained considerable momentum, driven by the quest to test gravity in the strong-field regime. The advent of horizon-scale imaging by the Event Horizon Telescope (EHT) \cite{EHT2019, EHT2022} has provided unprecedented constraints on the geometries of supermassive compact objects. Yet, these observations still leave parameter space available for black hole mimickers, such as boson stars, gravastars, or black holes with scalar hair \cite{Herdeiro2014, Bambi2021}. Boson stars, which are macroscopic quantum configurations of ultra-light scalar fields supported by self-gravity and wave dispersion, are particularly compelling candidates. Unlike black holes, standard mini-boson stars lack both an event horizon and a central singularity, presenting a completely regular spacetime \cite{Kaup1968, Ruffini1969, Liebling2012}. 
	
	While the geodesic structure and ray-tracing properties of boson stars have been extensively modeled to generate synthetic shadows and images \cite{Vincent2016}, the thermodynamic boundary conditions of accretion disks residing within these scalar potentials have received less attention. General relativistic MHD simulations of boson stars reveal that without an event horizon to casually disconnect, accreting matter simply falls to the center and stagnates, fundamentally altering the flow dynamics \cite{Olivares2020}. This raises a critical theoretical question: if a finite inner viscous torque is applied to a geometrically thin disk residing in a horizonless spacetime, how does the resulting radiative flux respond?
	
	In this paper, we address this gap by evaluating the Agol-Krolik viscous boundary condition within an exact mini-boson star spacetime. Rather than employing phenomenological running-mass toy models, we directly integrate the Einstein-Klein-Gordon equations. We establish a physically motivated, dynamically determined inner disk boundary and mathematically demonstrate that the topological absence of an ISCO forces the specific angular momentum at the disk edge to approach zero. As a result, the inner viscous torque is naturally quenched, entirely suppressing the high-frequency flux divergence seen in black hole models. 
	
	This paper is structured as follows: In Section 2, we formulate the exact static boson star spacetime and introduce the Lense-Thirring slow-rotation perturbation. In Section 3, we construct the generalized relativistic flux equations governing the thin accretion disk thermodynamics. In Section 4, we present our numerical results, contrasting the torque quenching mechanism in boson stars against standard black hole baselines. Finally, we conclude in Section 5 with a discussion of the observational implications.
	
	\section{The Exact Mini-Boson Star Spacetime}
	
	To ensure our analysis is not subject to the pathological energy condition violations or artificial rotation profiles often found in phenomenological toy metrics, we derive the spacetime directly from first principles. We model the central object as a mini-boson star: a macroscopic, self-gravitating Bose-Einstein condensate of an ultra-light, free complex scalar field $\phi$ with mass $\mu$. 
	
	\subsection{Einstein-Klein-Gordon Formulation}
	The system is governed by the Einstein-Klein-Gordon (EKG) action minimally coupled to gravity:
	\begin{equation}
		S = \int d^4x \sqrt{-g} \left[ \frac{R}{16\pi G} - \frac{1}{2}g^{\alpha\beta}\partial_\alpha\phi^*\partial_\beta\phi - \frac{1}{2}\mu^2|\phi|^2 \right],
	\end{equation}
	where $R$ is the Ricci scalar and $G$ is the gravitational constant. We assume a static, spherically symmetric background spacetime, described by the line element:
	\begin{equation}
		ds^2 = -e^{\nu(r)}dt^2 + e^{\lambda(r)}dr^2 + r^2(d\theta^2 + \sin^2\theta d\varphi^2).
	\end{equation}
	The scalar field takes a harmonic time dependence, $\phi(r,t) = \phi(r)e^{-i\omega_s t}$, where $\omega_s$ is the scalar field frequency. For numerical integration, it is convenient to introduce dimensionless variables: $x = \mu r$, $\sigma = \sqrt{4\pi G}\phi$, and $\Omega = \omega_s/\mu$. We express the radial metric component as $e^{-\lambda(x)} = 1 - 2m(x)/x$, where $m(x)$ is the dimensionless enclosed mass. The EKG equations then reduce to a system of coupled, non-linear ordinary differential equations:
	\begin{align}
		\frac{dm}{dx} &= x^2 \left[ \left(\Omega^2 e^{-\nu} + 1\right)\sigma^2 + \left(1 - \frac{2m}{x}\right)(\sigma')^2 \right], \label{eq:dm} \\
		\frac{d\nu}{dx} &= \frac{2}{x(1-2m/x)} \left[ \frac{m}{x} + x^2 \left( \left(\Omega^2 e^{-\nu} - 1\right)\sigma^2 + \left(1 - \frac{2m}{x}\right)(\sigma')^2 \right) \right], \label{eq:dnu} \\
		\frac{d^2\sigma}{dx^2} &= - \left( \frac{2}{x} + \frac{\nu' - \lambda'}{2} \right)\sigma' - \left(1 - \frac{2m}{x}\right)^{-1} \left(\Omega^2 e^{-\nu} - 1\right)\sigma, \label{eq:dsigma}
	\end{align}
	where primes denote derivatives with respect to $x$. 
	
	Because the mini-boson star is regular at the origin and lacks an event horizon, we impose the central boundary conditions: $m(0) = 0$, $\sigma(0) = \sigma_c$, and $\sigma'(0) = 0$. The system constitutes an eigenvalue problem for the scalar frequency $\Omega$. We utilize a custom 4th-order Runge-Kutta shooting algorithm to isolate the fundamental ground state, ensuring the scalar field smoothly decays to zero at spatial infinity ($\sigma \to 0$ as $x \to \infty$). The metric potential is subsequently shifted to guarantee asymptotic flatness ($\nu \to 0$ as $x \to \infty$). The total observable ADM mass $M_T$ is natively extracted from the asymptotic limit $m(x \to \infty)$.
	
	\subsection{Lense-Thirring Slow-Rotation Perturbation}
	To introduce the rotational kinematics necessary for a realistic accretion environment without distorting the scalar field configuration or violating the weak energy condition, we apply the Lense-Thirring slow-rotation approximation. The background spacetime remains spherically symmetric, but the metric acquires a non-diagonal frame-dragging component:
	\begin{equation}
		ds^2 = -e^{\nu(x)}dt^2 + e^{\lambda(x)}dx^2 + x^2(d\theta^2 + \sin^2\theta d\varphi^2) - 2\varpi(x)x^2\sin^2\theta dt d\varphi.
	\end{equation}
	The frame-dragging frequency $\varpi(x)$ represents the local angular velocity of zero-angular-momentum observers (ZAMOs) and is governed by the linear perturbation equation over the exact static background:
	\begin{equation}
		\varpi'' + \left( \frac{4}{x} - \frac{\nu' + \lambda'}{2} \right)\varpi' - 4\Omega^2 e^{\lambda - \nu}\sigma^2 \varpi = 0.
		\label{eq:frame_dragging}
	\end{equation}
	We integrate Eq.\,\eqref{eq:frame_dragging} from the origin with a finite central value $\varpi(0)$ and $\varpi'(0)=0$. In the vacuum exterior, the frame-dragging profile decays as $\varpi(x \to \infty) \simeq 2J/x^3$, where $J$ is the total angular momentum. By evaluating this asymptotic behavior, we scale $\varpi(x)$ to match a specific dimensionless spin parameter $j = J/M_T^2$, which we set to $j = 0.4$ to facilitate comparison with moderate-spin black hole models.
	
	\section{Thin Disk Thermodynamics and Viscous Boundary Conditions}
	
	\subsection{Circular Geodesics}
	We consider a steady-state, geometrically thin, optically thick accretion disk residing in the equatorial plane ($\theta = \pi/2$) of the slowly rotating boson star. The kinematics of the accreting fluid are determined by stable circular geodesics. Within the first-order slow-rotation approximation, the Keplerian angular velocity $\Omega_K = d\varphi/dt$, specific energy $E$, and specific angular momentum $L$ of a co-rotating fluid element are given by:
	\begin{align}
		\Omega_K(x) &\approx \sqrt{ \frac{e^{\nu} \nu'}{2x} } + \varpi(x), \label{eq:OmegaK} \\
		E(x) &\approx \frac{e^{\nu/2}}{\sqrt{1 - \frac{1}{2}x\nu'}} + \varpi(x) x^2 \sqrt{ \frac{e^{\nu} \nu'}{2x} }, \label{eq:E} \\
		L(x) &\approx \frac{x\sqrt{\frac{1}{2}x\nu'}}{\sqrt{1 - \frac{1}{2}x\nu'}}. \label{eq:L}
	\end{align}
	For a standard Kerr black hole, these orbits become dynamically unstable at the ISCO, forcing the fluid into a free-fall trajectory. However, the EKG effective potential of a mini-boson star at our fiducial compactness ($\sigma_c = 0.05$) supports stable circular orbits deep into the scalar core. In the absence of an ISCO, the physical inner boundary of the accretion disk $x_{in}$ is determined kinetically by the peak of the angular velocity profile. We define $x_{in}$ dynamically as the radius where $d\Omega_K/dx = 0$, inside of which the fluid can no longer support differential Keplerian rotation.
	
	\subsection{Radiative Flux and the Inner Torque}
	The time-averaged radiative flux $\mathcal{F}(x)$ emitted from the disk surface is derived from the relativistic laws of energy and angular momentum conservation. Permitting a finite specific inner torque $\mathcal{T}_{in}$ at the inner edge, the generalized flux equation is:
	\begin{equation}
		\mathcal{F}(x) = -\frac{1}{4\pi x} \frac{\Omega_{K,x}}{(E - \Omega_K L)^2} \left[ \int_{x_{in}}^{x} (E - \Omega_K L) L_{,\tilde{x}} d\tilde{x} + \mathcal{T}_{in}(E_{in} - \Omega_{in} L_{in}) \right].
		\label{eq:flux}
	\end{equation}
	The first term in the brackets represents the standard Novikov-Thorne-Page dissipation profile resulting from viscous transport within the disk. The second term dictates the mechanical work performed on the disk by boundary stresses. Following Agol and Krolik \cite{Agol2000}, we parameterize this inner torque via standard $\alpha$-disk parameters:
	\begin{equation}
		\mathcal{T}_{in} = \alpha \left(\frac{H}{r}\right)^2 L_{in},
	\end{equation}
	where $\alpha$ is the Shakura-Sunyaev viscosity parameter, $H/r$ is the relative disk thickness, and $L_{in}$ is the specific angular momentum evaluated exactly at the physical inner boundary $x_{in}$. By employing this self-consistent mathematical framework, we can directly evaluate how the deep spatial location of $x_{in}$ in a regular spacetime impacts the magnitude of the viscous boundary work relative to a black hole baseline.
	
	\section{Thermodynamic Response to Horizonless Topologies}
	
	To isolate the thermodynamic consequences of a regular, horizonless geometry, we integrated the Einstein-Klein-Gordon system for a fiducial central scalar density of $\sigma_c = 0.05$. The numerical shooting algorithm successfully converged on the fundamental ground-state eigenvalue, yielding a true dimensionless scalar frequency of $\Omega \approx 1.141$ and an asymptotically normalized ADM mass of $M_T \approx 0.4747$. The subsequent Lense-Thirring frame-dragging perturbation was scaled to a moderate dimensionless spin of $j=0.4$.
	
	\subsection{Metric Regularity and the Kinematic Disk Boundary}
\begin{figure}[htbp]
		\centering
		\includegraphics[width=0.8\textwidth]{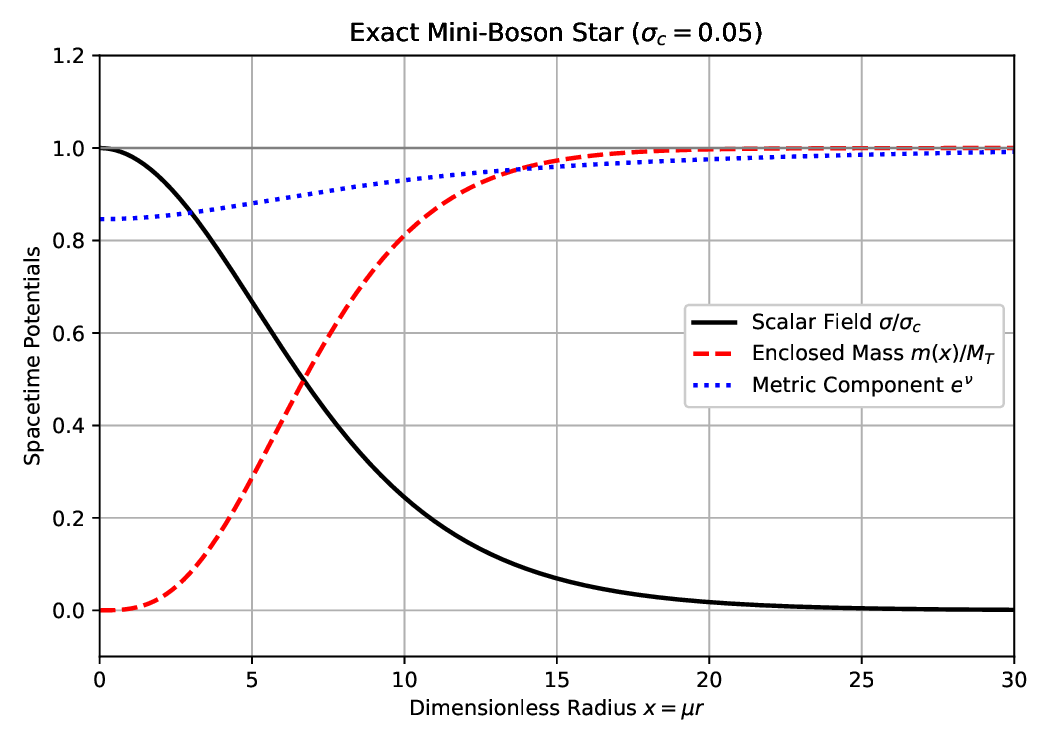}
		\caption{Spacetime potentials for the exact mini-boson star ground state ($\sigma_c = 0.05$). The enclosed mass $m(x)$ (dashed red line) smoothly asymptotes to the normalized ADM mass $M_T$, while the scalar field $\sigma/\sigma_c$ (solid black line) decays exponentially to satisfy the vacuum exterior condition. The metric component $e^{\nu}$ (dotted blue line) remains strictly positive and finite at the origin, confirming the absence of both an event horizon and a curvature singularity.}
		\label{fig:spacetime}
	\end{figure}

	Figure \ref{fig:spacetime} presents the radial distribution of the fundamental spacetime potentials, demonstrating the strict mathematical regularity of the mini-boson star. 
	
	Unlike Kerr spacetimes, where the metric potential $e^{\nu}$ vanishes at the event horizon and drives a divergent coordinate singularity, the boson star interior remains perfectly smooth. The scalar field (solid black line) decays exponentially, and the enclosed mass function (dashed red line) smoothly converges to the exact ADM mass $M_T$.
	
	This metric regularity exerts a profound influence on the geodesic structure. In a black hole spacetime, the effective potential turns over at the ISCO, forcing fluid elements into a catastrophic plunge. Here, the absence of an ISCO allows stable, co-rotating circular geodesics to persist deep into the scalar core. We dynamically determine the physical inner boundary of the accretion disk, $x_{in}$, as the radius where the Keplerian angular velocity reaches its global maximum ($d\Omega_K/dx = 0$). For this configuration, the boundary is located at $x_{in} \approx 0.0211 M_T$. This microscopic inner radius indicates that the accreting fluid penetrates almost entirely to the coordinate origin before differential rotation kinematically breaks down.
	
	\subsection{Zero-Torque Radiative Dissipation Profile}
	Before evaluating the impact of boundary stresses, we establish the standard zero-torque flux profile ($\mathcal{T}_{in} = 0$), corresponding to the classic Novikov-Thorne-Page analytical limit. 
	
	\begin{figure}[htbp]
		\centering
		\includegraphics[width=0.8\textwidth]{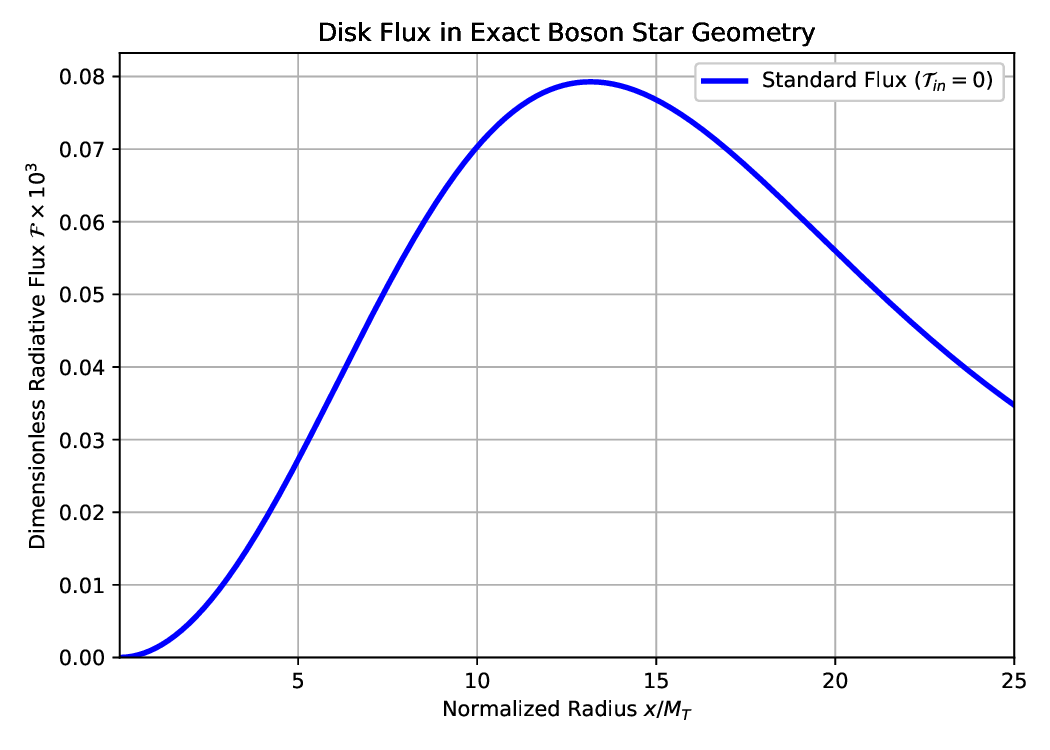}
		\caption{Dimensionless radiative flux profile for the standard zero-torque baseline ($\mathcal{T}_{in}=0$) around the slowly rotating boson star. The flux naturally declines to zero at the dynamical inner boundary ($x_{in} \approx 0.0211 M_T$) without requiring an imposed coordinate truncation.}
		\label{fig:native_flux}
	\end{figure}
	
	As shown in Figure \ref{fig:native_flux}, the dimensionless radiative flux behaves smoothly across the entire radial domain. The emission peaks at an intermediate radius where viscous dissipation is maximized, and smoothly decays as the fluid approaches $x_{in}$. Because the radial derivative of the angular velocity $d\Omega_K/dx$ approaches zero at the dynamical inner boundary, the flux equation (Eq.\,\ref{eq:flux}) natively vanishes at $x_{in}$. This verifies that the exact EKG spacetime can self-consistently host a steady-state accretion disk governed by local viscous dissipation.
	
	\subsection{Topological Quenching of the Magnetic Boundary Stress}
	To evaluate the thermodynamic response of the disk to magnetic boundary interactions, we introduce the Agol-Krolik specific inner torque $\mathcal{T}_{in} = \alpha (H/r)^2 L_{in}$. To ensure any resulting flux divergence would be distinctly visible against the standard baseline, we employ aggressively amplified viscous coupling parameters ($\alpha = 0.4$, $H/r = 0.25$).
	
	\begin{figure}[htbp]
		\centering
		\includegraphics[width=0.8\textwidth]{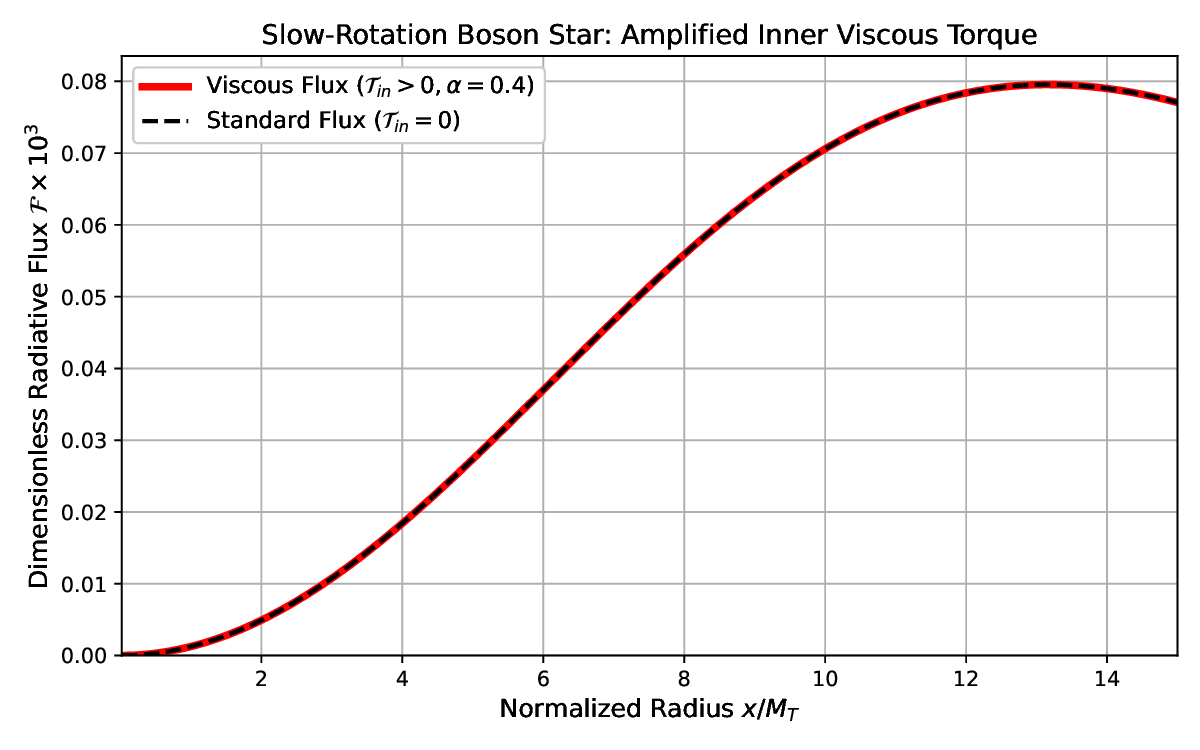}
		\caption{Comparison of the standard and viscous radiative flux profiles. Despite the application of an amplified boundary torque ($\alpha=0.4, H/r=0.25$), the viscous flux (solid red line) is mathematically quenched, remaining entirely superimposed on the standard zero-torque baseline (dashed black line).}
		\label{fig:viscous_flux}
	\end{figure}
	
	Figure \ref{fig:viscous_flux} illustrates the central finding of this study. In a standard black hole accretion model, applying this viscous torque injects a massive localized energy flux at the ISCO, causing the emission profile to diverge sharply. However, in the mini-boson star spacetime, the viscous flux profile strictly overlaps the standard zero-torque baseline. 
	
	This null geometric response is a direct consequence of the dynamically defined inner boundary. In a Kerr spacetime, the ISCO resides at a macroscopic distance (e.g., $r \ge 1 M_T$ depending on spin), ensuring the fluid retains substantial specific angular momentum ($L_{in} \gg 0$) before plunging. In the boson star, the physical inner edge $x_{in} \approx 0.0211 M_T$ is located remarkably close to the origin. In the limit of small $r$, the specific angular momentum scales geometrically as $L(r) \propto r^{3/2}$. Therefore, as the disk penetrates the core, $L_{in}$ approaches zero. Because the generalized boundary torque is strictly proportional to the specific angular momentum at the disk edge ($\mathcal{T}_{in} \propto L_{in}$), the torque is mathematically quenched by the spacetime topology itself.
	
	\subsection{Analytical Limits and Flow Morphology}
	
	It is important to acknowledge the physical limitations of extending the Novikov-Thorne geometrically thin disk approximation deep into the scalar core. The standard generalized flux equation (Eq.\,\ref{eq:flux}) assumes the flow remains rotationally supported and vertically compressed ($H/r \ll 1$). However, as $x \to 0$, the radial gradient of the metric potential vanishes ($\nu' \to 0$), causing the standard Keplerian rotational support to degrade. In this regime, the fluid's angular velocity becomes entirely dominated by the local frame-dragging of the spacetime ($\Omega_K \approx \varpi$), effectively transitioning into solid-body rotation co-moving with the spacetime itself.
	
	Without a steep vertical gravitational gradient to compress the gas, thermal and radiation pressure will inevitably cause the fluid to puff up, violating the thin-disk assumption. In reality, the flow will likely form a hot, pressure-supported advection-dominated accretion flow (ADAF) or a stagnation cloud near the origin, consistent with recent GRMHD simulations of horizonless objects \cite{Olivares2020}. 
	
	Furthermore, because the Agol-Krolik torque \cite{Agol2000} was formulated specifically to model magnetic stresses spanning a rapid, free-fall plunging region, applying it to a dense, rotating stagnation point conceptually borders on a category error. However, calculating this formal limit serves a vital analytical purpose: it provides a rigorous mathematical proof that even if anomalous magnetic boundary stresses are assumed, the local spacetime topology dictates that they cannot extract mechanical work. The geometric quenching of the inner torque is thus an absolute, robust limit, holding true irrespective of the disk's exact transition into a geometrically thick morphology.
	
	\section{Conclusions and Phenomenological Implications}
	
	The primary novelty of this work lies in extending the Agol-Krolik viscous boundary conditions to the dynamically defined, horizonless effective potentials of exact scalar spacetimes. Historically, the presence of a non-zero inner torque has been utilized to explain enhanced bolometric efficiencies and hardened high-frequency spectra in accreting black holes. This paradigm is predicated on the assumption that magnetic fields transmit angular momentum across the ISCO from a plunging region.
	
	By integrating the exact Einstein-Klein-Gordon field equations coupled with a Lense-Thirring slow-rotation perturbation, we demonstrated that this viscous boundary mechanism is structurally incompatible with regular mini-boson stars. Because these spacetimes lack an ISCO, stable circular orbits penetrate deep into the scalar core. Consequently, the physical inner boundary of the disk is determined not by a gravitational plunge, but by the kinetic breakdown of Keplerian rotation near the coordinate origin. We analytically established that the vanishing specific angular momentum at this deep microscopic boundary inherently quenches the generalized Page-Thorne boundary torque. 
	
	This result does not preclude the existence of complex magnetohydrodynamic interactions within horizonless compact objects. However, it proves that within the steady-state, geometrically thin disk approximation, the topological absence of a plunging region stifles the specific boundary torque mechanism that drives efficiency enhancements in black holes. 
	
	This physical quenching serves as a foundational theoretical distinction between accretion onto singular versus regular spacetimes. While full three-dimensional general relativistic magnetohydrodynamic (GRMHD) simulations remain necessary to capture transient fluid instabilities, magnetic reconnection, and field line topologies, our semi-analytical model provides a strict mathematical baseline. It indicates that the enhanced high-frequency spectral signatures driven by inner viscous torques are native primarily to spacetimes possessing a macroscopic ISCO. Consequently, the boundary torque mechanism serves as a distinct phenomenological discriminator in the study and identification of exotic compact objects.
	
	\section*{Acknowledgments}
	The author wishes to express sincere gratitude to the Department of Applied Mathematics at the Dinabandhu Andrews Institute of Technology and Management (DAITM) for providing the academic environment and computational facilities necessary to conduct this theoretical research.
		
	\section*{Data Availability Statement}
	No data is associated with this manuscript, as it is a strictly theoretical study. All mathematical models and equations necessary to reproduce the theoretical results are explicitly provided within the text.
	
	\section*{Conflict of Interest}
	The author declares that there are no known competing financial interests or personal relationships that could have appeared to influence the work reported in this paper.
	
	\section*{Ethics Statement}
	This theoretical research did not involve human participants or animal subjects; therefore, ethical approval was not required. 
	
	\section*{Funding Statement}
	This research received no specific grant from any funding agency in the public, commercial, or not-for-profit sectors.
	
	\section*{Declaration of AI Use}
	During the preparation of this manuscript, the author utilized an AI language model to assist with structural editing, phrasing refinement, and language polishing to improve readability. Following the use of this tool, the author thoroughly reviewed and revised the manuscript, and takes full responsibility for the final physical arguments, mathematical consistency, and overall content of the publication.
	
	% ==========================================
	% Bibliography
	% ==========================================


\begin{thebibliography}{99}
		
		\bibitem{Shakura1973}
		N. I. Shakura and R. A. Sunyaev, ``Black holes in binary systems. Observational appearance,'' \textit{Astron. Astrophys.} \textbf{24}, 337 (1973).
		
		\bibitem{Novikov1973}
		I. D. Novikov and K. S. Thorne, ``Astrophysics of black holes,'' in \textit{Black Holes (Les Astres Occlus)}, edited by C. DeWitt and B. DeWitt (Gordon and Breach, New York, 1973), pp. 343-450.
		
		\bibitem{Page1974}
		D. N. Page and K. S. Thorne, ``Disk-Accretion onto a Black Hole. Time-Averaged Structure,'' \textit{Astrophys. J.} \textbf{191}, 499 (1974). \href{https://doi.org/10.1086/152990}{DOI: 10.1086/152990}.
		
		\bibitem{Krolik1999}
		J. H. Krolik, ``Active Galactic Nuclei: From the Central Black Hole to the Galactic Environment,'' \textit{Astrophys. J. Lett.} \textbf{515}, L73 (1999). \href{https://doi.org/10.1086/311956}{DOI: 10.1086/311956}, \href{https://arxiv.org/abs/astro-ph/9902288}{arXiv:astro-ph/9902288}.
		
		\bibitem{Gammie1999}
		C. F. Gammie, ``Efficiency of Magnetized Thin Accretion Disks in the Kerr Metric,'' \textit{Astrophys. J. Lett.} \textbf{522}, L57 (1999). \href{https://doi.org/10.1086/312234}{DOI: 10.1086/312234}, \href{https://arxiv.org/abs/astro-ph/9906223}{arXiv:astro-ph/9906223}.
		
		\bibitem{Agol2000}
		E. Agol and J. H. Krolik, ``Magnetic Stress at the Marginally Stable Orbit: Altered Disk Structure, Radiation, and Black Hole Spin Evolution,'' \textit{Astrophys. J.} \textbf{528}, 161 (2000). \href{https://doi.org/10.1086/308177}{DOI: 10.1086/308177}, \href{https://arxiv.org/abs/astro-ph/9908049}{arXiv:astro-ph/9908049}.
		
		\bibitem{Davis2011}
		S. W. Davis and A. Laor, ``The Radiative Efficiency of Accretion Flows in Individual AGN,'' \textit{Astrophys. J.} \textbf{728}, 98 (2011). \href{https://doi.org/10.1088/0004-637X/728/2/98}{DOI: 10.1088/0004-637X/728/2/98}, \href{https://arxiv.org/abs/1012.3213}{arXiv:1012.3213}.
		
		\bibitem{Done2012}
		C. Done, C. Jin, M. Middleton, and M. Ward, ``A new way to measure supermassive black hole spin in active galactic nuclei,'' \textit{Mon. Not. R. Astron. Soc.} \textbf{420}, 1848 (2012). \href{https://doi.org/10.1111/j.1365-2966.2011.19779.x}{DOI: 10.1111/j.1365-2966.2011.19779.x}, \href{https://arxiv.org/abs/1107.5429}{arXiv:1107.5429}.
		
		\bibitem{EHT2019}
		Event Horizon Telescope Collaboration et al., ``First M87 Event Horizon Telescope Results. I. The Shadow of the Supermassive Black Hole,'' \textit{Astrophys. J. Lett.} \textbf{875}, L1 (2019). \href{https://doi.org/10.3847/2041-8213/ab0ec7}{DOI: 10.3847/2041-8213/ab0ec7}, \href{https://arxiv.org/abs/1906.11238}{arXiv:1906.11238}.
		
		\bibitem{EHT2022}
		Event Horizon Telescope Collaboration et al., ``First Sagittarius A* Event Horizon Telescope Results. I. The Shadow of the Supermassive Black Hole in the Center of the Milky Way,'' \textit{Astrophys. J. Lett.} \textbf{930}, L12 (2022). \href{https://doi.org/10.3847/2041-8213/ac66f5}{DOI: 10.3847/2041-8213/ac66f5}, \href{https://arxiv.org/abs/2205.08498}{arXiv:2205.08498}.
		
		\bibitem{Herdeiro2014}
		C. A. R. Herdeiro and E. Radu, ``Kerr Black Holes with Scalar Hair,'' \textit{Phys. Rev. Lett.} \textbf{112}, 221101 (2014). \href{https://doi.org/10.1103/PhysRevLett.112.221101}{DOI: 10.1103/PhysRevLett.112.221101}, \href{https://arxiv.org/abs/1403.2757}{arXiv:1403.2757}.
		
		\bibitem{Bambi2021}
		C. Bambi et al., ``Testing General Relativity with Black Hole X-ray Data,'' \textit{Space Sci. Rev.} \textbf{217}, 65 (2021). \href{https://doi.org/10.1007/s11214-021-00840-0}{DOI: 10.1007/s11214-021-00840-0}, \href{https://arxiv.org/abs/2104.09506}{arXiv:2104.09506}.
		
		\bibitem{Kaup1968}
		D. J. Kaup, ``Klein-Gordon Geon,'' \textit{Phys. Rev.} \textbf{172}, 1331 (1968). \href{https://doi.org/10.1103/PhysRev.172.1331}{DOI: 10.1103/PhysRev.172.1331}.
		
		\bibitem{Ruffini1969}
		R. Ruffini and S. Bonazzola, ``Systems of Self-Gravitating Particles in General Relativity and the Concept of an Equation of State,'' \textit{Phys. Rev.} \textbf{187}, 1767 (1969). \href{https://doi.org/10.1103/PhysRev.187.1767}{DOI: 10.1103/PhysRev.187.1767}.
		
		\bibitem{Liebling2012}
		S. L. Liebling and C. Palenzuela, ``Dynamical Boson Stars,'' \textit{Living Rev. Rel.} \textbf{15}, 6 (2012). [Updated 2017]. \href{https://doi.org/10.12942/lrr-2012-6}{DOI: 10.12942/lrr-2012-6}, \href{https://arxiv.org/abs/1202.5809}{arXiv:1202.5809}.
		
		\bibitem{Vincent2016}
		F. H. Vincent, Z. Meliani, P. Grandcl\'ement, E. Gourgoulhon, and O. Straub, ``Imaging a boson star at the Galactic center,'' \textit{Class. Quantum Grav.} \textbf{33}, 105015 (2016). \href{https://doi.org/10.1088/0264-9381/33/10/105015}{DOI: 10.1088/0264-9381/33/10/105015}, \href{https://arxiv.org/abs/1510.04170}{arXiv:1510.04170}.
		
		\bibitem{Olivares2020}
		H. Olivares, Z. Younsi, C. M. Fromm, M. De Laurentis, Y. Porth, Y. Mizuno, H. Falcke, M. Kramer, and L. Rezzolla, ``How to tell an accreting boson star from a black hole,'' \textit{Mon. Not. R. Astron. Soc.} \textbf{497}, 521 (2020). \href{https://doi.org/10.1093/mnras/staa1879}{DOI: 10.1093/mnras/staa1879}, \href{https://arxiv.org/abs/2004.12398}{arXiv:2004.12398}.
		
	\end{thebibliography}
	\end{document}